\documentclass[aps,prl,floatfix,twocolumn,amsfonts,amssymb,amsmath]{revtex4-1} 
\usepackage{dsfont} 
\usepackage{float} 
\usepackage[colorlinks=true,linkcolor=black,citecolor=black,filecolor=black,urlcolor=black,breaklinks=true]{hyperref}
\usepackage{mathtools}
\usepackage[capitalize]{cleveref}
\usepackage{physics}
\usepackage{tikz}
\usepackage{xcolor}
\usepackage{csquotes}
\usepackage{multibib} 

\newcommand{\m}{\mathrm}
\newcommand{\Ito}{It\^o}
\newcommand{\ml}{\mathcal{L}}
\newcommand{\mv}{\mathcal{V}}
\newcommand{\my}{\mathcal{Y}}
\newcommand{\mq}{\mathcal{Q}}
\newcommand{\diag}{\mathrm{diag}} 
\newcommand{\sgn}{\mathrm{sgn}}
\newcommand{\fs}[1]{\textcolor{black}{#1}}

\DeclarePairedDelimiter\superket{\lvert}{\rangle\negthinspace\negthinspace\rangle}

\DeclarePairedDelimiter\superbra{\langle\negthinspace\negthinspace\langle}{\rvert} 

\DeclarePairedDelimiterX\inproduct[2]{\langle\negthinspace\negthinspace\langle}{\rangle\negthinspace\negthinspace\rangle}{#1 \delimsize\vert #2}

\DeclarePairedDelimiter\lcom{\text{[\negthinspace[}}{\text{]\negthinspace]}}

\DeclarePairedDelimiterX\superdyad[2]{\lvert}{\rvert}{#1 \delimsize\rangle\negthinspace\negthinspace\rangle \negthinspace\langle\negthinspace\negthinspace\langle #2}

\newcommand{\ketflip}[1]{\mathop{\superket{#1}}}
\newcommand{\braflip}[1]{\mathop{\superbra{#1}}}
\newcommand{\braketflip}[2]{\mathop{\inproduct{#1}{#2}}}

\begin{document}

\title{Noise-Induced Quantum Synchronization}
\author{Finn Schmolke}
\author{Eric Lutz}
\affiliation{Institute for Theoretical Physics I, University of Stuttgart, D-70550 Stuttgart, Germany}

\begin{abstract}
    Synchronization is a widespread phenomenon in science and technology. We here study noise-induced synchronization in a  quantum spin chain subjected to local Gaussian white noise. We demonstrate   stable (anti)synchronization between the endpoint magnetizations of a quantum $XY$ model with transverse field of arbitrary length. Remarkably, we show that noise applied to a single spin suffices to reach stable (anti)synchronization, and find that the two synchronized end spins are entangled. We additionally determine the optimal noise amplitude  that leads  to the fastest synchronization along the chain, and further compare the optimal synchronization speed  to the fundamental Lieb-Robinson bound for information propagation.
    \end{abstract}
 
\maketitle 

Synchronization is ubiquitous in classical nonlinear systems. Self-sustained periodic oscillators swing in unison, and are thus synchronized, when their phases (frequencies) are locked \cite{ble88,boc02,pik03,ace05,mos02,ani07,bal09}. Synchronous behavior plays  a central role in many interconnected systems in   fields ranging  from biology and chemistry  to  physics and engineering. Different mechanisms of classical synchronization have been identified \cite{ble88,boc02,pik03,ace05,mos02,ani07,bal09}. For instance, forced synchronization may  be generated by an external drive, whereas spontaneous synchronization may be induced by the mutual coupling between subsystems in the absence of  external forcing. Another intriguing effect is noise-induced synchronization \cite{pik84,zho02,ter04,gol05,gol05a,nak07,ter08,lai11}, which has recently been observed in  sensory neurons \cite{nei02} and in lasers \cite{sun14}. 

In the past years, the study of synchronization  has been extended to the quantum domain \cite{goy06,zhi08,hei11,gio12,lud13,mar13,gio13,lee13,lee14,wal14,hus15,wal15,lor17,rou18,rou18a,son18,dav18,cab19,kop19,kar20,tyn20}. Quantum synchronization has been examined in systems with a classical analogue, such as nonlinear van der Pol oscillators, as well as in systems without a classical counterpart, such as qubits \cite{goy06,zhi08,hei11,gio12,lud13,mar13,gio13,lee13,lee14,wal14,hus15,wal15,lor17,rou18,rou18a,son18,dav18,cab19,kop19,kar20,tyn20}.  Both forced and spontaneous synchronizations have been investigated in the quantum regime \cite{goy06,zhi08,hei11,gio12,lud13,mar13,gio13,lee13,lee14,wal14,hus15,wal15,lor17,rou18,rou18a,son18,dav18,cab19,kop19,kar20,tyn20}. Quantum entrainment may starkly differ from classical entrainment: it has indeed been shown to exhibit counterintuitive nonclassical features, such as   enhanced synchronization of far-detuned oscillators and suppressed  synchronization of resonant oscillators \cite{lor17}. Quantum synchronization  has recently been experimentally observed in spin-1 systems \cite{kop20,las20}. 

We here investigate noise-induced synchronization in an isolated quantum many-body  system. By locally applying Gaussian white noise to a quantum spin chain of arbitrary length, we show that stable (anti)synchronization  between local spin observables may be  achieved when a given condition, on the length of the chain and on the sites at which noise is applied, is satisfied. In that case, local spin observables oscillate with the same frequency, a dynamical criterion for quantum synchronization that has been widely applied \cite{gio12,gio13,cab19,kar20,tyn20,gio19}. Remarkably, stable (anti)synchronization can be established between the two ends of the chain,  even when noise is applied  to only a single site in between. While noise is often assumed to be detrimental for quantum features owing to decoherence, we  establish that the  synchronized end spins are entangled by evaluating their concurrence \cite{woo98}. We finally analyze the time needed to fully synchronize the two ends of the  chain as a function of the noise strength and of the length of the chain,  when noise is added close to one end. We find the existence of an optimal noise amplitude that leads to the shortest  synchronization time (or fastest synchronization rate). This optimal time scales like the cube of the chain length, thus stronger than the linear dependence given by the  Lieb-Robinson bound, which provides a fundamental upper limit on the speed of  information propagation in a quantum system \cite{lie72}. 

\begin{figure*}[t]
	\centering
	\begin{tikzpicture}
	\node (a) [label={[label distance=-.5 cm]136: \textbf{a)}}] at (0,0) {\includegraphics{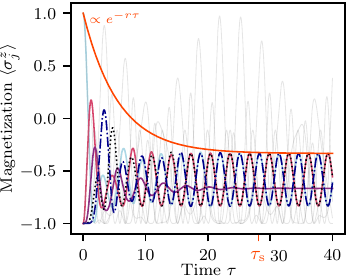}};	\node (a) [label={[label distance=-.5 cm]136: \textbf{b)}}] at (5.7,0) {\includegraphics{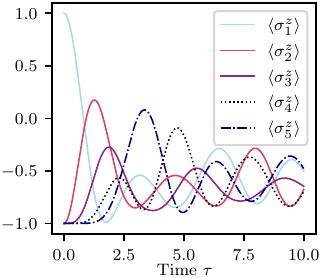}}; \node (a) [label={[label distance=-.5 cm]136: \textbf{c)}}] at (11.15,0) {\includegraphics{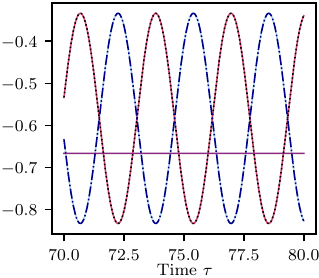}};
	\end{tikzpicture}
	\caption{Stable synchronization. a) Evolution of the local  magnetizations   $\langle \sigma^z_j\rangle$ of the quantum $XY$ spin chain, \cref{eq:Spin-chain_Hamiltonian}, of length $N=5$ with  white noise amplitude $\gamma =0.2$ applied to site $u=3$ (grey lines in the background show the corresponding noise-free evolutions).  The stable synchronization condition \eqref{9} is obeyed. The system has $\# \Lambda = \lfloor 5^2/4 \rfloor = 6$ eigenmodes with respective decay constants $m_{12} = m_{14} = m_{23} = 2/3$, $m_{13} = 4/9$, $m_{15} = 8/9$ and $m_{24}=0$. The smallest {nonzero} one, $r=\gamma m_{13}$, sets the decay to the synchronized state \eqref{eq:stable_mode} (orange line). b) No synchronization occurs for times shorter than the synchronization time $\tau_\m s=5/r$. c) Stable synchronization between  the end spins, $\langle \sigma^z_1 \rangle$ and $\langle \sigma^z_5 \rangle$, as well as between $\langle \sigma^z_2 \rangle$ and $\langle \sigma^z_4 \rangle$, appears for times larger than $\tau_\m s$.  The initial state is $\ket{\Psi(0)} = \ket{1}_1 \otimes \bigotimes_{j=2}^N \ket{0}_j$, where  ($\ket{0}_j$, $\ket{1}_j$) are the  ground and excited states of qubit $j$.} \label{f1}
\end{figure*}

\textit{Synchronization model.} We consider an isolated quantum many-particle system with Hamiltonian $H_0$ subjected to a stochastic perturbation of the form $\xi(t) V$, where $\xi(t)$ describes classical noise that couples to operator $V$. For concreteness and simplicity, we take a  quantum $XY$ chain of $N$ spins in a transverse field \cite{tak99}
\begin{equation}
    H_0 = \frac{J}{2}\sum_{j=1}^{{N-1}} (\sigma^x_j \sigma^x_{j+1} + \sigma^y_j \sigma^y_{j+1} )+ h \sum_{j=1}^N \sigma^z_j,
    \label{eq:Spin-chain_Hamiltonian}
\end{equation}
where $\sigma_j^{x,y,z}$ are the usual Pauli operators,  $J>0$ is the interaction parameter  and $h=1$ the field strength. We additionally choose delta-correlated   (white)
Gaussian noise,  $\langle \xi(t)\xi(t^\prime)\rangle = \Gamma \delta(t-t^\prime)$, with zero mean  and    amplitude $\Gamma$.
 We will in the following consider various  operators $V$, depending on the number and on the position of the sites the noise couples to. This many-particle system may be implemented using trapped ions  \cite{sch12,mon21}, where  noise is introduced by locally modulating ac-Stark shifts of the respective spin states \cite{mai19}. 

In order to examine the influence of noise on the quantum spin chain, and derive the synchronization condition, it is convenient to describe the time evolution of the system in Liouville space \cite{gya20}. In this formalism, a density matrix $\rho_\xi$ is mapped onto a vector $\ketflip{\rho_\xi}$
(often called supervector) in a higher-dimensional Hilbert space, and the von Neumann equation, $\dot \rho_\xi(t) = -i [H_0+\xi(t)V,\rho_\xi(t)]$, is transformed into the Schr\"odinger-like equation \cite{gya20}
\begin{align}
    \ketflip{\dot{\rho}_\xi(\tau)} = -i(\ml_0 + \xi(\tau)\mv)\ketflip{\rho_\xi(\tau)},
    \label{eq:Strato_SDE}
\end{align}
which can be analyzed using the usual tools of quantum mechanics. The Liouville superoperator $\ml_0$ is given by the supercommutator $\ml_0 = \lcom{H_0,\mathds{1}}/J = H_0/J \otimes \mathds{1} - \mathds{1} \otimes H_0^\m{T}/J$ and the perturbation superoperator by $\mv = \lcom{V,\mathds{1}}$. We have further introduced the normalized time $\tau = Jt$. Equation \eqref{eq:Strato_SDE} has the form of a stochastic differential equation with multiplicative noise (which we interpret using the Stratonovich convention) \cite{kam92}. Averaging over an ensemble of  noise realizations, Eq.~\eqref{eq:Strato_SDE} becomes \cite{sup}
\begin{align}
    \ketflip{\dot{\rho}(\tau)} = -(i\ml_0 + \gamma \mv^2/2) \ketflip{\rho(\tau)}\,,
    \label{eq:Lindbladian_in_Lspace}
\end{align}
where $\rho(\tau) = \langle \rho_\xi(\tau)\rangle$ is the averaged density operator and  $\gamma =\Gamma/J$ is  the reduced noise strength.

\begin{figure*}[t]
	\centering
\begin{tikzpicture}
	\node (a) [label={[label distance=-.7 cm]136: \textbf{a)}}] at (0,0) {\includegraphics{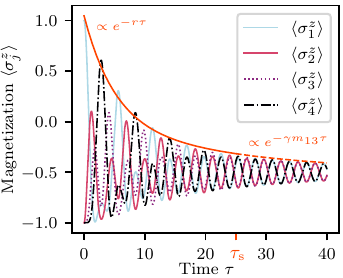}};	\node (a) [label={[label distance=-.7 cm]136: \textbf{b)}}] at (5.7,0) {\includegraphics{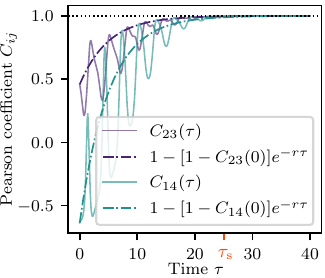}}; \node (a) [label={[label distance=-.7 cm]136: \textbf{c)}}] at (11.3,0) {\includegraphics{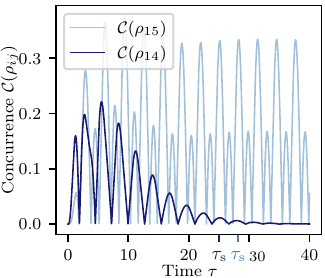}};
	\end{tikzpicture}	\caption{Transient synchronization and entanglement. a) Evolution of the local  magnetizations   $\langle \sigma^z_j\rangle$ of the quantum $XY$ spin chain, \cref{eq:Spin-chain_Hamiltonian}, of length $N=4$ with  white noise amplitude $\gamma =0.2$ applied to sites $u=2$ and $v=3$.  The stable synchronization condition \eqref{9} is not satisfied. The system has $\# \Lambda = \lfloor 4^2/4 \rfloor = 4$ eigenmodes with respective decay constants $m_{12} = m_{34} = m_{23}  = 1$ and $m_{13} = 1/5$. After  $\tau_\m s=5/(\gamma m_{13})$, transient synchronization (with decay rate {$r = \gamma m_{12}$}) appears.  b) Pearson correlation coefficients $C_{14}$ and $C_{23}$ showing transient synchronization  between the endpoints of the chain, $\langle \sigma^z_1 \rangle$ and $\langle \sigma^z_4 \rangle$, as well as between $\langle \sigma^z_2 \rangle$ and $\langle \sigma^z_3 \rangle$. c) The concurrence $C(\rho_{15})$ for stable synchronization (\cref{f1}a) displays nonzero steady oscillations after   $\tau_\m s$,  indicating entanglement between the edge spins, contrary to transient synchronization  $\mathcal{C}(\rho_{14})$ (\cref{f2}a). } \label{f2}
\end{figure*}

The stochastic perturbation  affects both  eigenmodes and  eigenfrequencies of the  unperturbed quantum system. We start from the spectral decomposition of the free evolution, $\ketflip{\rho_0(\tau)} = \exp(-i\ml_0\tau)\ketflip{\rho(0)}$,  given by
\begin{equation}
    \ketflip{\rho_0(\tau)} = \sum_{k,l} \exp(-i\Lambda_{kl}\tau) \ketflip{\nu_k,\nu_l} \braketflip{\nu_k,\nu_l}{\rho(0)},
    \label{eq:rho_in_normal_modes}
\end{equation}
where eigenfrequencies $\Lambda_{kl}$ and  eigenmodes $\ketflip{\nu_k,\nu_l}$ of the Liouvillian $\ml_0$ are related to the respective eigenvalues $\Lambda_k $ and eigenstates $\ket{\nu_k}$ of the Hamiltonian $H_0/J$ via  $\Lambda_{kl} = \Lambda_k - \Lambda_l$ and $\ketflip{\nu_k,\nu_l} = \ket{\nu_k} \otimes \ket{\nu_l}^\ast$ \cite{gya20}.
Eigenfrequencies always come in  pairs, $\Lambda_{kl} = -\Lambda_{lk}$. For weak noise ($\gamma\ll 1$),  eigenmodes and  eigenfrequencies of the perturbed system can be determined using  perturbation theory in Liouville space \cite{li14}. To first order, we obtain  
\begin{align}
\label{eq:perturbation_theory}
 \!\!\!  \Lambda^\m{p}_{kl} \simeq \Lambda_{kl} - i\gamma m_{kl}, \,\,\,
    {\ketflip{\nu_k,\nu_l}}^\m{p}\!\! \simeq {\ketflip{\nu_k,\nu_l}}^{(0)}\!\!\! - \gamma {\ketflip{\nu_k,\nu_l}}^{(1)},\!\!\!\!
\end{align}
where we have used the superscript $\m{p}$  to label the perturbed quantities \cite{com4}.

According to Eq.~\eqref{eq:perturbation_theory},  eigenmodes of the quantum many-body system  experience a selective exponential decay with rate $\gamma m_{kl}$. Stable synchronization occurs when all the modes decay to zero except one. This leads to a decoherence-free subspace \cite{lid98} with only a single eigenmode \cite{com1}: After a given time, which we  call synchronization time $\tau_{\m s}$, the system will  be in one eigenstate  ${\ketflip{\nu_k,\nu_l}}^\m{s}$ in Liouville space and oscillate with the corresponding eigenfrequency $\Lambda_{kl}^\m{s}$. On the other hand, transient synchronization appears when there is a clear timescale separation between the different decay times \cite{gio19}. In this situation,  a Liouville  eigenstate ${\ketflip{\nu_k,\nu_l}}^\m{t}$  which oscillates with frequency $\Lambda_{kl}^\m{t}$ can outlive all the others for a very long time---there is no synchronization otherwise \cite{sup}. Both types of synchronization occur in the quantum spin chain \eqref{eq:Spin-chain_Hamiltonian}.  The above route to synchronization is reminiscent of the classical synchronization mechanism known  as \enquote{suppression of natural dynamics} \cite{mos02,ani07,bal09}: beyond a critical forcing (coupling) amplitude in forced (spontaneous) synchronization, mode locking does not occur, but natural oscillations of the system are  suppressed, leaving it synchronized in a new mode \cite{ani92,pot01,bal02,hau06,usa14,wan21}. However,  the present   quantum phenomenon is  different: i) it is noise-induced, ii) it suppresses all the natural modes of the system except one, iii) it does not require a critical noise amplitude, and iv) it does not rely on  limit cycles.

\textit{Stable synchronization condition}. Let us now derive  the stable synchronization condition for the quantum spin chain \eqref{eq:Spin-chain_Hamiltonian}. We concretely focus on the local spin magnetizations,  $\langle \sigma^z_j \rangle=  \Tr[\sigma^z_j \rho(\tau)]$, at sites $j$.  Our first task is  to connect the abstract eigenstates of the Liouvillian (in Liouville space) to the physical eigenstates of the qubit (in Hilbert space). This is achieved by projecting the supervector $\ketflip{\rho(\tau)}$ onto the supervector $\ketflip{\sigma^z_j}$ \cite{sup}:
\begin{align}
    \sigma^z_j(\tau)
    = \braketflip{\sigma^z_j}{\rho(\tau)}
  =\sum_{kl} c_{kl} \exp(-i\tilde \Lambda_{kl} \tau) \epsilon_{j,kl}.
    \label{eq:local_energies}
\end{align}
The magnetization eigenmodes of the $j$th qubit are given by the projection $\epsilon_{j,kl} = {\braketflip{\sigma^z_j}{\nu_k,\nu_l}}$ and the coefficients $c_{kl} = {\braketflip{\nu_k,\nu_l}{\rho(0)}}$ depend on the initial excitations. {The magnetization frequencies $\tilde \Lambda_{kl}$ are thus a subset of $\{\Lambda_{kl}\}$.}

Our next task is to evaluate the decay constants $m_{kl}$. The quantum XY model \eqref{eq:Spin-chain_Hamiltonian} is integrable and can be  diagonalized exactly with the Jordan-Wigner transformation \cite{tak99}. 
The system has a total of $\#{\tilde \Lambda} = \lfloor N^2/4 \rfloor$ {magnetization} eigenfrequencies of which $\#{\tilde \Lambda}^\m{non} = \lfloor N/2 \rfloor$ are nondegenerate and $\#{\tilde \Lambda}^\m{deg} = \lfloor N^2/2(N/2-1) \rfloor$ are twofold degenerate (here, $\lfloor \cdot \rfloor$ denotes the floor function). The degeneracy strongly affects the decay rates. For nondegenerate eigenstates, the decay constants (in first-order perturbation theory) read $2m_{kl} = \braflip{\nu_k,\nu_l} \mv^2 \ketflip{\nu_k,\nu_l}$. On the other hand, since  the noise operator $V$ is real and symmetric in the Jordan-Wigner representation, the decay constants for degenerate eigenstates are $ 2m^\pm_{ab} = (\mv^2)_{aa} \pm |(\mv^2)_{ab}|$,   
where $\{\ketflip{a},\ketflip{b}\}$ denotes the degenerate eigenspace and $\fs{(\mv^2)_{ab}} = \braflip{a} \mv^2 \ketflip{b}$. The perturbation moreover lifts the degeneracy. The zeroth order eigenmodes are explicitly given by 
$\ketflip{a^\pm} =  (\ketflip{b} \pm \sgn\{\fs{(\mv^2)}_{ab}\}\ketflip{a})/{\sqrt{2}}$. In the Hilbert space of  $H_0$, we find that $\braketflip{\sigma^z_j}{a^\pm} = (1\pm\sgn\{\fs{(\mv^2)}_{ab}\})\epsilon_{j,a}/\sqrt{2}$.
Depending on the sign of $\fs{(\mv^2)}_{ab}$,  decay will therefore be either faster with $m^+_{ab}$ or slower with $m^-_{ab}$.
    
A. One-site noise. We proceed by  applying noise to  a single qubit of the chain located at site $u$ and set accordingly $V = \sigma^z_u$ \cite{com3}.
In this case, $\fs{(\mv^2)}_{ab} < 0$ and the rates are thus $m^-_{ab}$. We concretely obtain \cite{sup}
\begin{eqnarray}
	m^u_{kl}|_\m{non} &=& 
	\frac{{4}}{N+1} 
	\left[
		\sin(\frac{u k\pi}{N+1})^2+\sin(\frac{u l\pi}{N+1})^2
	\right]\nonumber \\
	&-&\frac{{16}}{(N+1)^2}
	\left[
		\sin(\frac{u k\pi}{N+1})^2\sin(\frac{u l\pi}{N+1})^2
	\right],
\label{totaldecay}
\end{eqnarray} 
for nondegenerate eigenstates, ${\tilde \Lambda}_{kl} \in {\tilde \Lambda}^\m{non}$. 
For degenerate eigenstates, ${\tilde \Lambda}_{kl} \in {\tilde \Lambda}^\m{deg}$, the second square bracket in Eq.~\eqref{totaldecay} is  multiplied by a factor 2.
The overall decay scales like $1/N$, implying slower decay for longer chains.

Stable synchronization is achieved when all the modes, except one, decay to zero. From Eq.~\eqref{totaldecay}, we find that $m^u_{kl}=0$ {for a single mode} is only possible when   \cite{sup}
\begin{align}
\label{9}
   \!\! \frac{N+1}{3} \in \mathbb{N}, \,\, 
    \frac{u}{3} \in \mathbb{N}, \,\,  {k} = \frac{(N+1)}{3}, \,\, {l} = 2k \,\,(\m{for}\, N \ge 5)
\end{align}
 is fulfilled \cite{com5}. 
The  synchronized mode is then  \cite{sup}
\begin{equation}
  \ket{{\epsilon}_{kl}}^\m{s}
    = {\frac{3}{N+1}} 
	\begin{dcases}
        \!\left(1,-1,0,-1,1,\ldots,1, -1\right)^\m{T}\!\!, \ N\text{even}\\
        \!\left(1,-1,0,-1,1,\ldots,-1, 1\right)^\m{T}\!\!, \ N\text{odd}.
    \end{dcases}
    \label{eq:stable_mode}
\end{equation}
with $\ket{\epsilon_{k l}} =(\epsilon_{1,kl},\epsilon_{2,kl}\dots, \epsilon_{N,kl})^\m{T}$. The corresponding eigenfrequency  is ${\Lambda^\m{s}_{kl}} = 2$. We see from \cref{eq:stable_mode} that the end magnetizations are synchronized (antisynchronized) for $N$ even (odd). Remarkably, noise at a single site suffices to (anti)synchronize  the endpoints of a  chain of arbitrary length. We note that the amplitude of the (anti)synchronized  mode scales inversely to the length.

\Cref{f1}a shows the time evolution of the magnetization $\langle \sigma^z_j \rangle$ (colored lines) for a chain of length $N=5$ and white noise applied to site $u=3$ (the grey lines in the background indicate the unperturbed evolution in the absence of noise for comparison). This case obeys the stable synchronization condition \eqref{9}. Oscillations are out-of-phase, and no synchronous behavior  is seen, for times smaller than the synchronization time $\tau_\m{s}$ (\cref{f1}b). However, for times larger than $\tau_\m{s}$, stable synchronization between the endpoints of the chain, $\langle \sigma^z_1 \rangle$ and $\langle \sigma^z_5 \rangle$, as well as between $\langle \sigma^z_2 \rangle$ and $\langle \sigma^z_3 \rangle$, appears (the magnetization $\langle \sigma^z_3 \rangle$ is independent of time in this regime) (\cref{f1}c).

B. Two-site noise. The effect of noise simultaneously applied to several sites may be studied in a similar manner.  For  two sites, $V= \sigma^z_u+  \sigma^z_v$, we find \cite{sup}
\begin{eqnarray}
    &&m^{u,v}_{kl}|_\m{non} =
    \frac{{4}}{N+1}
    \Bigg[
        \sin(\frac{uk\pi}{N+1})^2 
        + \sin(\frac{vk\pi}{N+1})^2 \nonumber\\
        &&+ \sin(\frac{ul\pi}{N+1})^2 
        + \sin(\frac{vl\pi}{N+1})^2\Bigg] \nonumber\\
        &&-\frac{{16}}{(N+1)^2}
        \Bigg[\sin(\frac{uk\pi}{N+1})^2
        + \sin(\frac{vk\pi}{N+1})^2\Bigg] \nonumber\\ 
        &&\times \Bigg[\sin(\frac{ul\pi}{N+1})^2 + \sin(\frac{vl\pi}{N+1})^2\Bigg],
\end{eqnarray}
for nondegenerate eigenfrequencies, and
\begin{eqnarray}
    m^{u,v}_{kl}|_\m{deg}\! &=& m^{u,v}_{kl}|_\m{non}\! - \!\frac{{16}}{(N+1)^2}\!\!
    \left[
        \sin(\frac{uk\pi}{N+1})\!\sin(\frac{ul\pi}{N+1}) \right.\nonumber\\
        &&+\left. \!\sin(\frac{vk\pi}{N+1})\!\sin(\frac{vl\pi}{N+1})
    \right]^2,
\end{eqnarray}
for degenerate eigenfrequencies. The {only} possible configuration such that $m^{u,v}_{k,l} = 0$ for a single  mode is \cite{sup}
\begin{align}
\label{12}
    \frac{u}{3} \in \mathbb{N}&,\qquad \frac{v}{3} \in \mathbb{N}, \qquad \frac{N+1}{3}\in \mathbb{N},\\
    k  &= \frac{N+1}{3}, \qquad  l = 2k,
\end{align}
These conditions are equivalent to those indicated in Eq.~\eqref{9} for a single-site noise.  They will thus lead to the same synchronized (antisynchronized) modes. The only difference is that the overall strength of the noise is here twice as large. The two time evolutions will hence be the same with the replacement $\gamma \to \gamma/2$.

\Cref{f2}a represents the dynamics of the magnetizations $\langle \sigma^z_j \rangle$  for a chain of length $N=4$ and white noise applied to sites $u=2$ and $v=3$. The stable synchronization condition \eqref{12} is here not satisfied. Yet, after the synchronization time $\tau_\m s$, transient synchronization is observed between the endpoints of the chain, $\langle \sigma^z_1 \rangle$ and $\langle \sigma^z_4 \rangle$, as well as between $\langle \sigma^z_2 \rangle$ and $\langle \sigma^z_3 \rangle$. The occurrence of (transient) in-phase oscillation between these qubits is further confirmed by the examination of   the corresponding Pearson correlation coefficients, defined as the ratio of the covariance and the respective standard deviations, $C_{ij}= \text{Cov}(\langle \sigma^z_i \rangle,\langle \sigma^z_j \rangle)/\sqrt{\text{Var}(\langle \sigma^z_j \rangle)\text{Var}(\langle \sigma^z_j \rangle)}$ \cite{bar89}. For  $\tau> \tau_\m s$, both $C_{14}(\tau)$ and $C_{23}(\tau)$ converge to one, implying maximum correlation,  and hence synchronous motion, between the local magnetizations (\cref{f2}b) \cite{com6}. We mention that the above quantum synchronization phenomena are robust against weak perturbations \cite{sup}. 

In order to analyze the entanglement properties of the synchronized edge spins, we  plot in \cref{f2}c the concurrence  {$\mathcal{C}(\rho_{ij}) = \m{max}\left(0,\sqrt{\kappa_1}-\sqrt{\kappa_2}-\sqrt{\kappa_3}-\sqrt{\kappa_3}\right)$}, where $\rho_{ij} = \Tr_{\{1,\ldots,n\}\setminus (i,j)}[\rho(t)]$ is the reduced density operator obtained by tracing out the rest of the chain, and {$\kappa_n$} are the ordered eigenvalues of  $\rho_{ij} \widetilde \rho_{ij}$, with  $\widetilde \rho_{ij}$   the spin flipped state \cite{woo98}. For stable synchronization (\cref{f1}a), ${\cal C}(\rho_{15})$ exhibit steady oscillations  with nonzero amplitude after $\tau_\m s$,  revealing that the two end spins are entangled despite the action of the  noise (the nondecaying mode is  insensitive to the external perturbation). By contrast, for transient synchronization (\cref{f2}a), ${\cal C}(\rho_{14})$ vanishes after $\tau_\m s$, and the corresponding spins are thus not entangled.

\begin{figure}[t]
    \centering
     \begin{tikzpicture}
	\node (a) [label={[label distance=-.5 cm]128: \textbf{a)}}] at (0,0) {\includegraphics{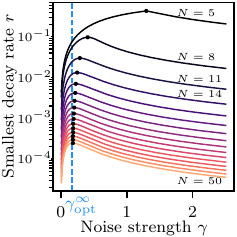}};	\node (a) [label={[label distance=-.5 cm]128: \textbf{b)}}] at (4.1,0) {\includegraphics{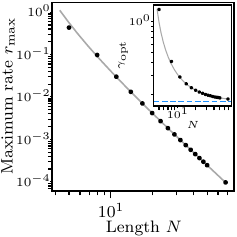}};
	\end{tikzpicture}
    \caption{Optimal synchronization. a) Decay rate $r$ as a function of the noise strength $\gamma$, for various  lengths $N$ {that fulfill the stable synchronization condition Eq.~\eqref{9}}. Dots  indicate the optimal   strength $\gamma_\text{opt}$ leading to fastest synchronization: $\gamma_\text{opt}$ tends to the nonzero  $\gamma_\text{opt}^\infty$ for large $N$ (blue dashed line). b)  Maximal  rate  $r_\m{max} = r(\gamma_\m{opt})$ as a function of $N$ (inset shows the corresponding  strength $\gamma_\m{opt}$).
   Grey lines are a fit with the function $f(N) = a + b/(N+c)^{\alpha}$ for $N > 5$ (main: $a = 0.00$, $b = 22.65$, $c = -1.75$, $\alpha = 2.94$; inset: $a = 0.17$, $b = 0.78$, $c = -4.84$, $\alpha = 1.01 $).
 }    
    \label{f3}
   
\end{figure}
\textit{Synchronization time.} The speed of signal propagation in discrete quantum systems with local interactions is upper bounded by the  Lieb-Robinson velocity \cite{lie72}, which in the XY model with transverse field is given by $v_\m{LR} = 2J$.
This finite group velocity defines an effective light cone beyond which the amount of transferred information decays exponentially.  Consequently, a minimal  time is needed for information to travel along a quantum spin chain. We here investigate the minimal time it takes  to fully (anti)synchronize  the two edges of the quantum XY model  \eqref{eq:Spin-chain_Hamiltonian} of arbitrary length $N$, as a function of the noise strength $\gamma$, and compare the result to the Lieb-Robinson bound. To that end, we consider single-site noise applied at site $u=3$, and  solve the quantum Liouville {equation \eqref{eq:Lindbladian_in_Lspace}} numerically for varying  $N$ and $\gamma$. We compute the  eigenvalues $-\mu_\alpha(N,\gamma) + i \lambda_\alpha(N,\gamma)$, with real part $\mu_\alpha(N,\gamma) \ge 0$ and imaginary part $\lambda_\alpha(N,\gamma)$. The eigenvalue with smallest real and nonvanishing imaginary part $\mu^\m s=\min\{\mu_\beta| \lambda_\beta \neq 0\}$ sets the decay of the synchronized mode. Consequently, $r(N,\gamma) = \m{min}\{\mu_\alpha > \mu^\m{s}| \lambda_\alpha \neq 0\}$, sets the relaxation time to the (anti)synchronized  state, as seen in \cref{f1,f2} (orange lines), as well as in \cref{f2}b for the Pearson coefficients (dashed lines). 
We thus define the synchronization time as $\tau_\m{s} = 5/r$.

\Cref{f3}a displays the decay rate $r$ as a function of the noise amplitude $\gamma$ for different   chain lengths $N$. We observe that  $r$ first sharply increases with increasing noise strength: intensifying small noise hence significantly speeds up the relaxation, and accordingly reduces  the synchronization time. However, beyond an optimal noise amplitude $\gamma_\m{opt}$, the decay rate progressively decreases, and the relaxation is slowed down. This slowing down is  related to   the phenomenon of noise-induced quantum  Zeno effect \cite{whi09,bur20}. The maximum decay rate $r_\m{max} = r(\gamma_\m{opt})$ scales as $1/N^3$, as seen in \cref{f3}b, in agreement with the scaling of the inverse gap of the Liouvillian of the XY model with boundary dissipation \cite{zni15}. The synchronization time $\tau_\m{s}$ therefore grows like the third power of  the number $N$ of lattice sites, indicating that bigger systems need longer to (anti)synchronize. This dependence is stronger than the $1/N$ scaling of decay rates set by the Lieb-Robinson bound  \cite{zni15,com5}. At the same time, the optimal noise strength $\gamma_\m{opt}$ decreases as $1/N$, like {the related} decay rates $m^u_{kl}$ in Eq.~\eqref{totaldecay}, before saturating at an asymptotic nonzero value $\gamma_\m{opt}^\infty$, independent of the length $N$ (blue dashed line). 

\textit{Conclusions.} We have demonstrated the occurrence of stable (anti)synchronization of the endpoints of an isolated quantum spin chain exposed to Gaussian white noise.  We have obtained  (equivalent) stable synchronization conditions, \cref{9} for one-site noise and \cref{12} for two-site noise, for this noise-induced phenomenon to happen in the quantum domain.  We have  additionally determined the optimal noise amplitude corresponding to the shortest synchronization time, and shown that the latter grows cubically with the system size, hence stronger than the linear Lieb-Robinson bound. Remarkably, noise applied at a single spin is enough to synchronize a  chain of arbitrary length and  synchronized edge spins are nonclassically correlated. This opens up the possibility to employ them for synchronization-based \cite{arg05,cho17} quantum communication systems.

\textit{Acknowledgements.}
We acknowledge support from the Vector Foundation and thank Kaonan Micadei for helpful discussions.

\clearpage
\widetext
\begin{center}
\textbf{\large Supplemental Material: Noise-Induced Synchronization in Quantum Systems}
\end{center}
\setcounter{equation}{0}
\setcounter{figure}{0}
\setcounter{table}{0} 
\setcounter{page}{1}
\makeatletter
\renewcommand{\theequation}{S\arabic{equation}}
\renewcommand{\thefigure}{S\arabic{figure}}
\renewcommand{\bibnumfmt}[1]{[S#1]}
\renewcommand{\citenumfont}[1]{S#1}


\renewcommand{\figurename}{Supplementary Figure}
\renewcommand{\theequation}{S\arabic{equation}}
\renewcommand{\thefigure}{S\arabic{figure}}
\renewcommand{\bibnumfmt}[1]{[S#1]}
\renewcommand{\citenumfont}[1]{S#1}
\renewcommand{\thesection}{\Roman{section}} 

The Supplemental Material contains details about (I) the calculation of the averaged  equation of motion in Liouville space, (II) the evaluation of the decay rates for the XY model, (III) the derivation of the condition for stable synchronization, (IV) an example of synchronization of two independent detuned subsystems coupled by white noise, (V) a demonstration of the robustness of synchronization against  weak perturbations, and (VI) the exact solution of a two-qubit system for which the Lieb-Robinson bound is attained.

\section{Averaged evolution equation in Liouville space}
\label{sec:mean_evolution}
We first derive the evolution equation of the averaged density operator $\rho(t)$ in Liouville space for a generic (possibly time dependent) quantum system subjected to white noise.
We start with a general Hamiltonian of the form 
\begin{align}
    H(t) = H_0(t) + \xi(t) V(t),
    \label{eq:general_Hamiltonian}
\end{align}
where the total Hamiltonian $H(t)$ consists of a non-fluctuating part $H_0(t)$ and a stochastic part $\xi(t) V(t)$, with $\xi(t)$ a delta-correlated Gaussian  noise with zero mean and amplitude $\Gamma$,
   $ \langle \xi(t)\rangle = 0$, $\langle \xi(t)\xi(t^\prime)\rangle = \Gamma \delta(t-t^\prime)$.
 For notational convenience, we drop the explicit time dependence of the Hamiltonian parts in the following. 

In Liouville space, the stochastic evolution of the superket $\ketflip{\rho_\xi(t)}$  is described by the Stratonovich stochastic differential equation \cite{gya20}
\begin{align}
    \ketflip{\dot{\rho}_\xi(t)} = -i(\ml_0 + \xi(t)\mv)\ketflip{\rho_\xi(t)} \qquad \text{(Stratonovich)},
    \label{eq:single_realization}
\end{align}
with the supercommutator for the \fs{non-fluctuating} part $\fs{\ml_0} = \lcom{H_0,\mathds{1}} = H_0 \otimes \mathds{1} - \mathds{1} \otimes H_0^\m{T}$ and the \fs{stochastic} part $\mv = \lcom{V,\mathds{1}}$.
In differential form this equation reads 
\begin{align}
    \dd{\vb{x}} = -i\ml_0 \vb{x} \dd{t} - i\sqrt{\Gamma}\mv \vb{x} \dd{W(t)} \qquad \text{(Stratonovich)},
    \label{eq:Kubo_Strat}
\end{align} 
where we have used the shorthand notation $\vb{x} = \ketflip{\rho_\xi(t)}$; $W(t)$ here denotes a Wiener process. \Cref{eq:Kubo_Strat} is of Langevin type with multiplicative noise and represents a generalized Kubo-oscillator model \cite{Kubo_1963,Kubo_1969}. We can hence identify the drift and diffusion coefficients for this process: the superket $\vb{x}$ undergoes Brownian motion in the complex plane with a constant drift given by $\ml_0$ and diffusion coefficient given by $\Gamma \mv^2$. As with the classical Kubo-oscillator, the noise thus leads to {phase diffusion}.

The Stratonovich equation \eqref{eq:Kubo_Strat} can be converted ito an equivalent \Ito\ equation describing the same process \cite{Gardiner_2004}
\begin{align}
    \dd{\vb{x}} = (-i\ml_0 - \Gamma \mv^2/2)\vb{x} \dd{t} -i\sqrt{\Gamma}\fs{\mv} \vb{x} \dd{W(t)} \qquad \text{(It\^o)}. 
    \label{eq:Kubo_Ito}
\end{align}
Transformation between \Ito\ and Stratonovich always comes at the expense of an additional drift term \cite{vanKampen_1981}, which here is given by half the diffusion coefficient. We exploit the properties of \Ito\ calculus to directly evaluate the average of \eqref{eq:Kubo_Ito}. By definition of the \Ito\ integral, $\vb{x}$ is always taken at the left endpoint of the interval, that is {before} a jump in $\fs{\dd{W(t)}}$. Thus, $\vb{x}$ and $\fs{\dd{W(t)}}$ are statistically independent and their average vanishes, $\langle \vb{x}\dd{W(t)}\rangle = 0$. We accordingly  arrive at  Eq.~(3) of the main text:
\begin{align}
    \ketflip{\dot{\rho}(t)} = -(i\ml_0 + \Gamma \mv^2/2) \ketflip{\rho(t)},
\end{align}
with $\ketflip{\rho(t)}= \langle \ketflip{\rho_\xi(t)} \rangle$. We note that the above Liouville space formalism  has also been recently employed to investigate  a noisy quantum Ising chain (with global instead of local coupling) in the context of quantum phase transitions \cite{dut16} and quantum simulations \cite{che17}.

\section{Computation of  the decay rates for the $XY$ model}
We next detail the computation of the decay rates for the $XY$ model with transverse field with Hamiltonian \cite{tak99}
\begin{align}
    H = J\sum_{j=1}^{N-1} (\sigma^+_j\sigma^-_{j+1} + \sigma^-_j\sigma^+_{j+1}) + h\sum_{j=1}^N \sigma^z_j\,.
\end{align}
We first perform a Jordan-Wigner transformation to reduce the dimensions, thereby mapping the interacting spin-1/2 systems to a chain of non-interacting spinless fermions \cite{tak99}. The transformed Hamiltonian is tridiagonal and reads
\begin{align}
    H_\m{JW} = J\sum_{j=1}^{N-1} (c^\dagger_j c_{j+1} + c^\dagger_{j+1} c_j) + h\sum_{j=1}^{\fs{N}} (2 c^\dagger_j c_j - \mathds{1}), 
\end{align}
where $c^\dagger_j$ and $c_j$ are the respective fermionic  ladder  operators. They are related to the usual Pauli operators by
\begin{align}
    c_j = \exp(i\pi\sum^{j-1}_{n=1} c^\dagger_nc_n) \sigma^-_j, \qquad c_j^{\dagger} = \exp(i\pi\sum^{j-1}_{n=1} c^\dagger_nc_n) \sigma^+_j.
\end{align}
In the $c_j$ basis, the evolution of the two-point correlations, $Z = \langle (c_j^\dagger c_k)_{1\le j, k \le N} \rangle$, follow a von Neumann-type equation
\begin{align}
    \dot{Z}(t) = i[\Omega,Z(t)],
\end{align} 
where $\Omega = \diag(J;2h;J)$ contains the interactions on the off-diagonals and the \fs{populations} on the diagonal.
The diagonal elements of $Z(t)$ therefore give the time evolution of the populations from which we obtain the magnetizations by simple rescaling, $\langle \sigma^z_j \rangle = 2Z_{jj}-1$.

Let us now apply Gaussian white noise $\xi(t)$ \fs{to} a single spin at site $u$ by choosing $V = \sigma^z_u$. In the Jordan-Wigner representation, the equation of motion then becomes 
\begin{align}
    \dot{Z}_\xi(t) = i [\Omega + 2\xi(t)Y,Z_\xi(t)],
\end{align}
with $Y = \dyad{\hat{\vb{e}}_u}$ in the canonical basis. 
In Liouville space, we accordingly have 
\begin{align}
    \ketflip{\dot{Z}_\xi(t)} = (i\mq +2\xi(t) \my)\ketflip{Z_\xi(t)},
    \label{eq:JW_single_realization}
\end{align}
with $\mq = \lcom{\Omega,\mathds{1}}$ and $\my = \lcom{Y,\mathds{1}}$.
Equation \eqref{eq:JW_single_realization} is of the same form as \eqref{eq:single_realization}. Averaging over the noise, we therefore find \begin{align}
    \ketflip{\dot{Z}(t)} = (i\mq - 2\Gamma\my^2)\ketflip{Z(t)},
    \label{eq:JW-Lspace-average}
\end{align}
with $\ketflip{Z(t)} =\langle \ketflip{Z_\xi(t)}\rangle$.
{The spin chain possesses an intrinsic time scale given by the interaction strength $J$. To arrive at statements independent of the specific value of $J$, we introduce the normalized time $\tau = Jt$.} 

Since \cref{eq:JW-Lspace-average} is of the form of a Schr\"odinger equation (with \enquote{Hamiltonian} $i\mq$ and \enquote{perturbation} $-\fs{2}\Gamma \my^2$), we can apply standard perturbation theory to evaluate the decay rates for small noise amplitude (with reduced strength $\gamma =\Gamma/J$). The  eigenvalues and eigenvectors of the tridiagonal Toepliz matrix $\Omega$ are respectively given by \cite{Reichel_2013}
\begin{align}
    \tilde \Lambda_k &= 2{h}/J + 2\cos(\frac{\pi k}{N+1}),\\
    \varphi_k &= \left(\sin(\frac{\pi k}{N+1}), \sin(\frac{2\pi k}{N+1}),\ldots,\sin(\frac{N\pi k}{N+1})\right)^\m{T} \sqrt{\frac{2}{N+1}}\,.
\end{align}
The (unperturbed) eigenfrequencies and normal modes of the noise-free system immediately follow as
\begin{align}
    \tilde{\Lambda}_{kl} &= 2\left(\cos(\frac{\pi k}{N+1}) - \cos(\frac{\pi l}{N+1})\right),
    \label{eq:eigval}\\
    \ketflip{\varphi_k,\varphi_l} &= \left(\sin(\frac{\pi k}{N+1})\sin(\frac{\pi l}{N+1}),\ldots,\sin(\frac{N \pi}{N+1})\sin(\frac{N \pi}{N+1}) \right)^\m{T} \frac{2}{N+1}.
    \label{eq:eval}
\end{align}
$\tilde\Lambda_{kl}$ and $\ketflip{\varphi_k,\varphi_l}$ are the eigenfrequencies and eigenmodes of the magnetizations $\langle \sigma^z_j \rangle$. Therefore the magnetization frequencies are a subset of the full set of frequencies $\{\tilde \Lambda_{kl}\} \subset \{\Lambda_{kl}\}$ and the magnetization modes are a subspace of the full eigensystem $\{\ketflip{\nu_k,\nu_l}\}$.
There are $\# \tilde \Lambda^\m{deg} = \lfloor N^2/(N/2-1) \rfloor$ twofold degenerate eigenfrequencies with 
\begin{align}
    \tilde \Lambda_{rs} = \tilde \Lambda_{N+1-s,N+1-r},
\end{align}
the others  are nondegenerate. For every degenerate eigenfrequency there are two orthonormal eigenvectors $\ketflip{a} = \ketflip{\varphi_r,\varphi_s}$ and $\ketflip{b} = \ketflip{\varphi_{N+1-s},\varphi_{N+1-r}}$.

We obtain the decay rates $m^u_{kl}$ of the nondegenerate frequencies by  computing the expectation  of the perturbation 
\begin{align}
    \fs{m^u_{kl}|_\m{non}} &= 2\braflip{\varphi_k,\varphi_l} \my^2 \ketflip{\varphi_k,\varphi_l},\\
             &= 2\braflip{\varphi_k,\varphi_l} \left(Y^2 \otimes \mathds{1} + \mathds{1} \otimes Y^2 - 2Y \otimes Y \right)\ketflip{\varphi_k,\varphi_l},\\
             &= 	\frac{4}{N+1} 
             \left[
                 \sin(\frac{uk\pi}{N+1})^2+\sin(\frac{ul\pi}{N+1})^2
             \right]-\frac{16}{(N+1)^2}
             \left[
                 \sin(\frac{uk\pi}{N+1})^2\sin(\frac{ul\pi}{N+1})^2
             \right],
    \label{eq:nondeg}
\end{align}
which is Eq.~(7) of the main text.

For the degenerate frequencies, we have to compute the eigenvalues and eigenvectors of the perturbation matrix 
\begin{align}
    P = 
    \begin{bmatrix}
       \fs{2(\my^2)_{aa}} & \fs{2(\my^2)_{ab}}\\
        \fs{2(\my^2)_{ba}} & \fs{2(\my^2)_{bb}}
    \end{bmatrix},
\end{align}
with $\fs{(\my^2)_{ab} = \braflip{a} \my^2\ketflip{b}}$.
Because $\Omega$ and $Y$ are real and symmetric, we have \fs{$(\my^2)_{ab} = (\my^2)_{ba}$} and we find \fs{$(\my^2)_{aa} = (\my^2)_{bb}$}. The decay rate and zeroth order eigenvector thus simplify to 
\begin{align}
    m^{\pm}_{ab} &= \fs{2(\my^2)_{aa} \pm 2|(\my^2)_{ab}|},
    \label{eq:mpm}\\
    v^+ &= \left(\frac{\fs{(\my^2)_{ab}}}{|\fs{(\my^2)_{ab}}|},1\right)^\m{T},
	\qquad
	v^- = \left(-\frac{\fs{(\my^2)_{ab}}}{|\fs{(\my^2)_{ab}}|},1\right)^\m{T}.
    \label{eq:vpm}
\end{align}
\fs{From \cref{eq:vpm} we can read off the perturbed eigenvectors. These are
\begin{align}
    \ketflip{a^\pm} = \frac{1}{\sqrt{2}}(\ketflip{a}\pm\sgn\{(\my^2)_{ab}\}\ketflip{a}).
\end{align}
The projection onto the magnetization eigenbasis produces the same magnetization eigenmode for $\ketflip{a}$ and
$\ketflip{b}$. That is, $\braketflip{\sigma^z_j}{a} = \braketflip{\sigma^z_j}{b}$.
We therefore finally obtain for the zeroth-order eigenmode
\begin{align}
    \braketflip{\sigma^z_j}{a^\pm} = \frac{1}{\sqrt{2}}(1\pm\sgn\{(\my^2)_{ab}\})\underbrace{\braketflip{\sigma^z_j}{a}}_{\mathclap{\equiv \epsilon_{j,a}, \text{ c.f. \cref{eq:magmode}}}}.
\end{align}
Depending on the sign of $(\my^2)_{ab}$ one of the corrections will always vanish while the other gains a factor of $\sqrt{2}$. The decay rate corresponding to the vanishing correction has to be discarded. In the case of single and two-site noise we always have $(\my^2)_{ab}<0$ and the decay rates are thus given by $m^-_{ab}$.}

\fs{Evaluating \cref{eq:mpm} explicitly}, we arrive at
\begin{equation}
    \begin{aligned}
        m^u_{kl} =& 
        \frac{4}{N+1} 
        \left[
            \sin(\frac{nk\pi}{N+1})^2+\sin(\frac{nl\pi}{N+1})^2
        \right]
        -\frac{16}{(N+1)^2}
        \left[
            \sin(\frac{nk\pi}{N+1})^2\sin(\frac{nl\pi}{N+1})^2
        \right] 
        \times 
        \begin{dcases}
            1, &\fs{\tilde \Lambda}_{kl} \in \tilde \Lambda^\m{non},\\
            2, &\fs{\tilde \Lambda}_{kl} \in \tilde \Lambda^\m{deg}
        \end{dcases}\,.
    \end{aligned}
    \label{sup:total_decay}
\end{equation} 
The case of two-site noise with  $V= \sigma^z_u+  \sigma^z_v$ can be treated similarly, yielding the decay rates 
\begin{eqnarray}
   m^{u,v}_{kl}|_\m{non} =&&
    \frac{{4}}{N+1}
    \left[
        \sin(\frac{uk\pi}{N+1})^2 + \sin(\frac{vk\pi}{N+1})^2 \right.
        \left.+ \sin(\frac{ul\pi}{N+1})^2 +\!\sin(\frac{vl\pi}{N+1})^2
    \right]   \nonumber\\
    -&&\frac{{16}}{(N+1)^2}
    \left[
        \sin(\frac{uk\pi}{N+1})^2 
    \right. 
   + \left.\sin(\frac{vk\pi}{N+1})^2\right] \times \left[
        \sin(\frac{ul\pi}{N+1})^2 + \sin(\frac{vl\pi}{N+1})^2
    \right],
\end{eqnarray}
for nondegenerate eigenfrequencies, and
\begin{eqnarray}
    m^{u,v}_{kl}|_\m{deg} &=& m^{u,v}_{kl}|_\m{non}\! - \!\frac{{16}}{(N+1)^2}
    \left[
        \sin(\frac{uk\pi}{N+1})\sin(\frac{ul\pi}{N+1}) \right.
        +\left. \sin(\frac{vk\pi}{N+1})\sin(\frac{vl\pi}{N+1})
    \right]^2,
\end{eqnarray}
for degenerate eigenfrequencies. The {only} possible configuration such that $m^{u,v}_{k,l} = 0$ for a single  mode is 
\begin{align}
    u = 3p, \,\,\,\, v = 3q, \,\,\,\, \frac{N+1}{3}\in \mathbb{N}, \,\,\,\, k = \frac{N+1}{3},\,\,\,\, l = 2k,
\end{align}
for $p,q \in [1,\lfloor N/3 \rfloor]$. These conditions are equivalent to those indicated in Eq.~(8) of the main text for a single-site noise\fs{, which will be derived in the next section}.  They will thus lead to the same (anti)synchronized  modes. The only difference is that the overall strength of the noise is here twice as large. The two time evolutions will hence be the same with the replacement $\gamma \to \gamma/2$.

The perturbative evaluation of the decay rates holds for small noise strength, $\gamma \ll 1$, corresponding to the linear regime in Fig.~3a of the main text. The mode with the smallest decay rate may change for different values of $\gamma$, and hence synchronization depends on $\gamma$ in general. However, the stable synchronization conditions (Eqs.~(9) and (12) of the main text)  ensure that synchronization takes place in a decoherence-free subspace, which remains unaffected by the noise for all values of $\gamma$. We thus have stable synchronization even in the quantum Zeno regime (Fig.~3a), as long as the smallest decay rate $r$ remains finite, that is, the synchronization time $\tau_\m{s}$ does not diverge.

\section{Stable synchronization condition}
In this Section, we derive the stable synchronization condition given by Eq.~(8) of the main text. To this end, we try to find a configuration $l,k,N,u$ for which every mode except a single one decays. We require that only a single pair $(k,l)$ fulfills $m^u_{kl} = 0$. By examining \cref{sup:total_decay}, we find that this is achieved when
\begin{align}
	\sin(\frac{uk\pi}{N+1}) \overset{!}{=} 0 \land \sin(\frac{ul\pi}{N+1}) \overset{!}{=} 0.
	\label{eq:sin0}
\end{align} 
However, if $k$ solves the above equation, every integer multiple of $k$  is also a solution (the same holds for $l$ of course). A sufficient condition is  given by $uk = N+1$. We further require that $ul = N+1$, but we are also constrained by $k \neq l$. As result, this condition becomes $ul = 2uk = N+1$. 

The configuration $k,2k$ is the one with the smallest possible $k,l$ that solves \cref{eq:sin0}. Because the value for $k$ cannot exceed the length of the chain $k \in [1,N]$, we additionally have 
\begin{align}
	k &< N,\\
	l = 2k &< N,\\
	3k &> N.
	\label{eq:req}
\end{align}
These conditions ensure that \cref{eq:sin0} has the unique solution $k,2k$. The next value to solve \cref{eq:sin0} would be $3k \notin [1,N]$, but this case is excluded by the third condition \eqref{eq:req}.

The remaining task is to find $u$ and $N$. From \eqref{eq:req} it follows that $(N+1)/3 = k$. Hence, $N+1$ needs to be divisible by $3$. Finally, we obtain $uk = u(N+1)/3 = N+1 \Rightarrow u = 3$. Putting everything together, we eventually find a configuration that achieves persistent oscillations with a single mode if 
\begin{align}
	N+1	\ \text{is divisible by }3, 
	\qquad k = \frac{N+1}{3},
	\qquad l = \frac{2(N+1)}{3},
	\qquad u = 3\,.
  \label{eq:single_decay_conditions}
\end{align}
The same effect also occurs for larger values of $u$ as long as it is less than half of the length of the chain, that is,  if $u < \lceil N/2\rceil$ is fulfilled. With increasing $N$, more and more values for $u$ become possible, where $u = 3,6,9,\ldots$. Note that if $u$ exceeds $\lceil N/2\rceil$ then, by symmetry of the problem, this would have the same outcome as if we had chosen $u^\prime = N+1-u$. 

Finally, we explicitly determine the synchronized mode. The magnetization of the $j$th spin evolves according to 
\begin{align}
    \sigma^z_j(\tau) 
    = \braketflip{\sigma^z_j}{\rho(\tau)}
    = \sum_k c_{kk}\epsilon_{j,kk} + \sum_{k\neq l} c_{kl} \exp(-i\tilde \Lambda_{kl}\tau) \epsilon_{j,kl},
    \label{eq:magmode}
\end{align}
where $\epsilon_{j,kl} \equiv \braketflip{\sigma^z_j}{\nu_k,\nu_l} = 2\braketflip{\hat{\vb{e}}_j,\hat{\vb{e}}_j}{\varphi_k,\varphi_l}$ ($k \neq l$) is the projection of the Liouville normal mode onto the magnetization subspace of the local spin $\sigma^z_j$.
To obtain the magnetization modes, we build a $N$-dimensional vector containing the local magnetization $\sigma^z_j$ of the $j$th spin in the $j$th component
\begin{align}
    \ket{\epsilon} 
    = 
    \begin{bmatrix}
        \sigma^z_1(\tau)\\
        \sigma^z_2(\tau)\\
        \vdots\\
        \sigma^z_N(\tau)
    \end{bmatrix}
    = \sum_k c_{kk}\epsilon_{j,kk} + \sum_{k \neq l}
    c_{kl} e^{-i\tilde \Lambda_{kl}\tau}\underbrace{
    \begin{bmatrix}
        \epsilon_{1,kl}\\
        \epsilon_{2,kl}\\
        \vdots\\
        \epsilon_{N,kl}
    \end{bmatrix}}_{= \ket{\epsilon_{kl}}}
    = \sum_k c_{kk}\epsilon_{j,kk} + \sum_{k\neq l} c_{kl} \exp(-i\tilde \Lambda_{kl}\tau) \underbrace{\sum_j \epsilon_{j,kl} \hat{\vb{e}}_j}_{= \ket{\epsilon_{kl}}}\,.
\end{align}
The magnetization mode corresponding to the eigenfrequency $\tilde \Lambda_{kl}$ is thus given by $\ket{\epsilon_{kl}}$. In the Jordan--Wigner representation the magnetization modes can be conveniently calculated by 
\begin{align}
    \ket{\epsilon_{kl}} = 2\sum_j \braketflip{\hat{\vb{e}}_j,\hat{\vb{e}}_j}{\varphi_k,\varphi_l} \hat{\vb{e}}_j = \sum_j \varphi_k^j\varphi_l^j \hat{\vb{e}}_j, \quad k\neq l
\end{align}
since $\sigma^z_j(\tau) = 2\braketflip{\hat{\vb{e}}_j,\hat{\vb{e}}_j}{Z} - 1$.

Independently of the specific value of $u$, we obtain the non-decaying mode ${\ket{\epsilon_{kl}}^\m{s}}$ by inserting {$k = (N+1)/3$ and $l = 2(N+1)/3$} into \cref{eq:eval}:
\begin{align}
	{\varphi_k^\m{s}} 
	&= \left(\sin(\frac{\pi}{3}),\sin(2\frac{\pi}{3}),\ldots,\sin(N\frac{\pi}{3})\right)^\m{T}\sqrt{\frac{2}{N+1}},\\
	{\varphi_l^\m{s}} 
	&= \left(\sin(\frac{\pi}{3}),\sin(2\frac{\pi}{3}),\ldots,\sin(N\frac{\pi}{3})\right)^\m{T}\sqrt{\frac{2}{N+1}},
\end{align}
In the Jordan--Wigner subspace, we then explicitly obtain
\begin{align}
	{\ket{\epsilon_{kl}}^\m{s}} 
    &= 2\sum_j \braketflip{\hat{\vb{e}}_j,\hat{\vb{e}}_j}{\varphi_k^\m{s},\varphi_l^\m{s}} \hat{\vb{e}}_j
    = \\ 
	&\frac{{4}}{N+1} \bigg(\sin(\frac{\pi}{3})\sin(2\frac{\pi}{3}),\sin(2\frac{\pi}{3})\sin(4\frac{\pi}{3}),\sin(\pi)\sin(2\pi),\ldots,\sin(N\frac{\pi}{3})\sin(2N\frac{\pi}{3})\bigg)^\m{T},
	\label{eq:single_mode}\\
	&= \frac{3}{4}\frac{4}{N+1}
	\begin{dcases}
		& \left(1,-1,0,1,-1,\ldots,1, -1\right)^\m{T}, \qquad N = \m{even}, \\
		& \left(1,-1,0,1,-1,\ldots,-1, 1\right)^\m{T}, \qquad N = \m{odd}\,.
	\end{dcases}
  \label{eq:shape_of_the_single_mode}
\end{align}
In the stable mode, \cref{eq:shape_of_the_single_mode}, every third spin has zero amplitude and adjacent spins in between are in anti-phase. Due to the noise, the chain breaks up into \fs{$(N+1)/3$} two-qubit systems which are separated by single equilibrated qubits. 
The corresponding eigenfrequency is $\tilde \Lambda^\m{s}_{kl} = 2$ (see \cref{eq:eigval}).

\begin{figure}[t]
    \centering
    \begin{tikzpicture}
    \node (a) [label={[label distance=-.5 cm]136: \textbf{a)}}] at (0,0) {\includegraphics{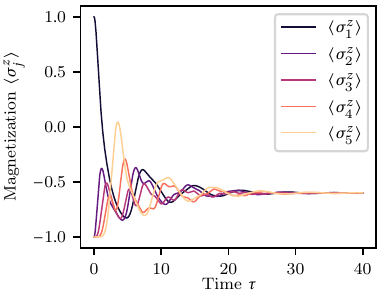}};	\node (a) [label={[label distance=-.5 cm]136: \textbf{b)}}] at (6.5,0) {\includegraphics{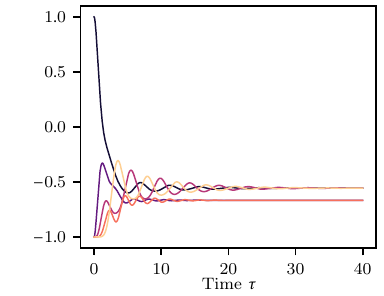}};
    \end{tikzpicture}
    \caption{Violation of the synchronization condition. Evolution of the five-qubit quantum $XY$ Ising chain with white noise applied to (a) the first site $V = \sigma^z_1$ and to (b) the second site $V = \sigma^z_2$ (because the chain is centrosymmetric, the same dynamics is also seen for either $V = \sigma^z_{1(2)}$ or $V = \sigma^z_{4(3)}$). No synchronization occurs in both these cases. This is the generic behavior of the noisy spin chain. The reduced noise strength is $\gamma = 1$.}
    \label{fig:nosync}
\end{figure}
When the conditions for stable (or transient) synchronization are not obeyed, we observe damped evolution which results in a nonoscillating (asynchronous) steady state. This corresponds to the generic behavior of the randomly perturbed Ising chain. We illustrate this  situation with the five-qubit example of the main text, applying white noise to either the first site, $V = \sigma^z_1$, (Fig.~\ref{fig:nosync}a) or to the second site, $V = \sigma^z_2$, (Fig.~\ref{fig:nosync}b). The same behavior appears in an arbitrary   Ising chain configuration $(N,u)$ with single site white noise applied at random sites (not satisfying the synchronization conditions).

\section{Noise-induced synchronization of independent subsystems}
\begin{figure}[t]
    \centering
    \begin{tikzpicture}
    \node (a) [label={[label distance=-.5 cm]136: \textbf{a)}}] at (0,0) {\includegraphics{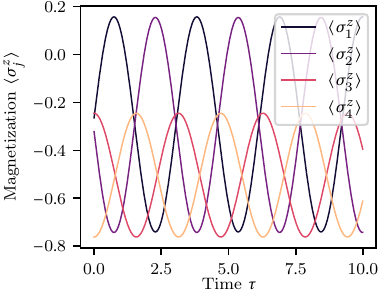}};	\node (a) [label={[label distance=-.5 cm]136: \textbf{b)}}] at (6.5,0) {\includegraphics{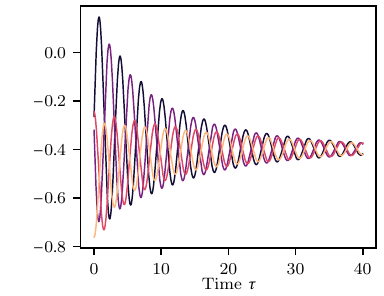}};
    \end{tikzpicture}
    \caption{Transient synchronization of  two independent two-qubit subsystems $A$ (with spins $\langle\sigma^z_{1(2)}\rangle$) and $B$ (with spins $\langle\sigma^z_{3(4)}\rangle$) coupled through the white noise operator $V = (\sigma^+_2\sigma^-_3 + \sigma^-_2\sigma^+_3)$, applied locally to the qubits 2 and 3 (Eq.~\ref{s41}). a) The local magnetizations $\langle \sigma^z_j\rangle$ initially oscillate out of phase with (slightly) different frequencies. (b) The two subsystems $A$ and $B$  exhibit transient synchronization at longer times.}
    \label{fig:link-noise}
\end{figure}

Classical synchronization requires a notion of decomposability, that is, the possibility to separate a large system into several oscillating subsystems that synchronize when coupled \cite{pik03}. We show in this \fs{section} that the quantum Ising chain satisfies this property by partitioning it into two independent subchains. Coupling the latter two  by white noise will indeed induce (transient) synchronization. Let us, for concreteness, consider two separate two-qubit systems $A$ (qubits 1 and 2) and $B$ (qubits 3 and 4) with respective Hamiltonians
\begin{align}
\label{s41}
    H_{A(B)} = \omega_{1(3)}\sigma^z_{1(3)} + \omega_{2(4)}\sigma^z_{2(4)} + J(\sigma^+_{1(3)}\sigma^-_{2(4)} + \sigma^-_{1(3)}\sigma^+_{2(4)}),
\end{align}
The natural frequencies $\omega_i$ are chosen such that the  subsystem corresponding to $H_A$ has different normal frequencies than the subsystem corresponding  to $H_B$. The two subsystems are coupled via the white noise operator $V = (\sigma^+_2\sigma^-_3 + \sigma^-_2\sigma^+_3)$ applied locally to the qubits 2 and 3. The resulting dynamics is shown in \cref{fig:link-noise} (with $\omega_1 = 1.2$ and $\omega_2 = \omega_3 = \omega_4 = 1$). We observe that the local magnetizations initially oscillate out of phase (\cref{fig:link-noise}a) \fs{before transient synchronization sets in (\cref{fig:link-noise}b).}

\section{Robustness of synchronization against weak disturbances}

\begin{figure}[t]
    \centering
    \begin{tikzpicture}
    \node (a) [label={[label distance=-.5 cm]136: \textbf{a)}}] at (0,0) {\includegraphics{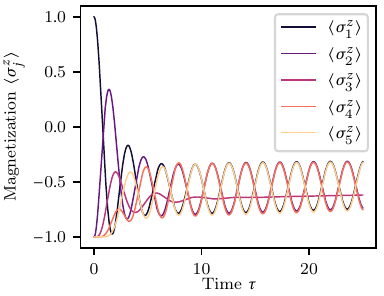}};	\node (a) [label={[label distance=-.5 cm]136: \textbf{b)}}] at (6.5,0) {\includegraphics{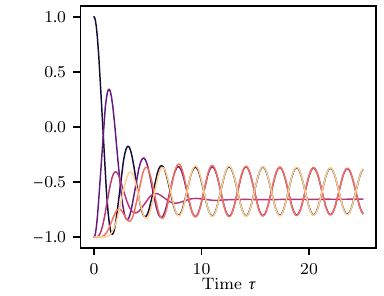}};
    \end{tikzpicture}
    \caption{Robustness of the synchronization effect against imperfections. In addition to the synchronization-inducing white noise  $V = \sigma^z_3$ at site 3, we apply another  operator $L$ at  site 1 of the five-qubit chain. (a) Single-site dissipator  $L = \sqrt{\lambda} \sigma^x_1$ and (b) single-site dephasor  $L = \sqrt{\lambda} \sigma^z_1$. The parameter $\lambda$ is chosen such that {$\lambda/\gamma = 0.01$}. Transient synchronization occurs for small disturbances, showing that the synchronization mechanism is robust to weak external perturbations.}
    \label{fig:robustness}
\end{figure}

We here investigate whether synchronization is robust against imperfections
such as single-site decay or dephasing on other sites.
To this end, we consider the five-qubit example of the main text and apply the single-site dissipator $L = \sqrt{\lambda} \sigma^x_1$ (which pumps energy into the system) and the single-site dephasing operator $L = \sqrt{\lambda} \sigma^z_1$ at site 1 (in addition to the synchronization-inducing white noise operator at site 3) \cite{bre02}. The results are presented in \cref{fig:robustness}. 
For small disturbances (we use {$\lambda/\gamma = 0.01$} in the plots),  synchronization becomes transient (the system finally reaches a steady state due to the perturbation), but remains robust. As a general rule, the quantum synchronization effect survives as long as the decoherence timescale (associated with dissipation or dephasing) is long enough so that the quantum properties of the system are preserved, as for all quantum technological applications \cite{bre02}.

\section{Two-qubit system achieves maximal relaxation rate}

\begin{figure}[t]
    \centering
    \begin{tikzpicture}
    \node (a) [label={[label distance=-.7 cm]136: \textbf{a)}}] at (0,0) {\includegraphics{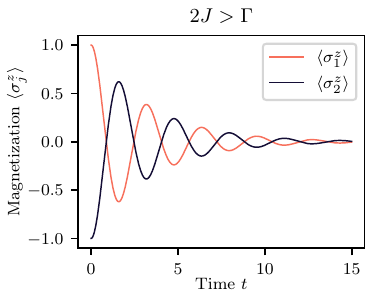}};	\node (a) [label={[label distance=-.7 cm]136: \textbf{b)}}] at (6,0) {\includegraphics{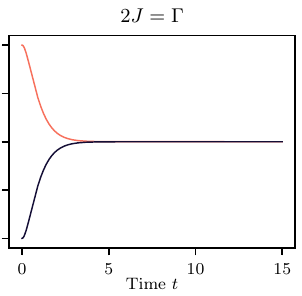}}; \node (a) [label={[label distance=-.7 cm]136: \textbf{c)}}] at (11.4,0) {\includegraphics{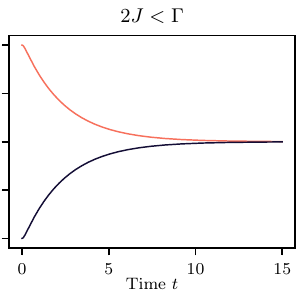}};
    \end{tikzpicture}
    \caption{Evolution of the local magnetizations \fs{$\langle \sigma^z_j \rangle$}, $j=1,2$, Eq.~\cref{eq:crit-damping}, of two qubits subject to centered white noise, Eq.~\cref{eq:H}. $\langle \bullet \rangle$ denotes the average over noise realizations. Three different regimes arise depending on the relative strength of interaction $J$ and noise $\Gamma$ (like for damped harmonic oscillators): a) Small damping  $\Gamma < 2J$, b) Critical damping  $\Gamma = 2J$, c) Strong damping  $\Gamma > 2J$. The qubits are initially in the ground and excited states, with respective populations, p1 = 1 and p2 = 0.}
    \label{fig:crit-damping} 
  \end{figure}

In this Section, we show  that a two-qubit system is able to relax to its steady state at the maximum Lieb-Robinson rate of $v_\m{LR} = 2J$ for noise acting on a single site (for noise acting on both sites, relaxation can happen faster because the disturbance has no distance to travel).
{For illustrative purposes, we do not use the normalized time $\tau$ here, but treat the problem with the original time variable $t$.}
The Hamiltonian is in this case
  \begin{align}
      H = \sigma^z_1(1 + \xi(t)) + \sigma^z_2 + J(\sigma^+_1 \sigma^-_2 + \m{h.c.}).
      \label{eq:H}
  \end{align}
We expect to find $\lfloor N^2/4 \rfloor=1$ normal mode. There can thus be no synchronization. Still, we take the opportunity to study the effect of noise in this simple system.  
To this end, we solve the corresponding master equation \cite{bre02}
  \begin{align}
      \dv{t} \rho(t) = i[H,\rho(t)] + \Gamma \mathcal{D}[\sigma^z_1] \rho(t),
  \end{align}
where $\mathcal{D}[A]\bullet = A \bullet A^\dagger - \frac{1}{2}\{A^\dagger A,\bullet\}$ denotes the dissipator.
Although the system contains too few qubits to exhibit synchronization, it still displays interesting behavior. For the local magnetizations, we indeed find
  \begin{align}
    \langle \sigma^z_{1(2)} \rangle 
    =& 
    p_1+p_2-1 \pm e^{-\Gamma t} \times\\
    &\begin{dcases}
      \frac{-4 J  + \Gamma (p_1-p_2)}{\sqrt{4 J^2-\Gamma^2}}\sin(\sqrt{4 J^2-\Gamma^2}t)+(p_1-p_2) \cos(\sqrt{4 J^2-\Gamma^2}t), & 2J > \Gamma,\\
      2Jt(p_1-p_2-2) + p_1-p_2 , & 2J = \Gamma,\\
      \frac{-4 J + \Gamma (p_1-p_2)}{\sqrt{\Gamma^2-4J^2}}\sinh(\sqrt{\Gamma^2-4J^2}t)+(p_1-p_2) \cosh(\sqrt{\Gamma^2-4J^2}t), & 2J < \Gamma\,.
    \end{dcases}
    \label{eq:crit-damping}
\end{align}

Interestingly, we obtain a similar evolution to that of a classical damped harmonic oscillator. Thus, the two-qubit system shows different oscillating behavior depending on the relative magnitude of the frequency $J$ and of the noise $\Gamma$.
The corresponding three regimes are displayed in \cref{fig:crit-damping}.
The case $2J > \Gamma$ leads to exponentially damped oscillations with a slightly reduced frequency. In the balanced case $2J = \Gamma$, equilibrium is reached most rapidly at the maximal rate $v_\m{LR}$ corresponding to the Lieb-Robinson bound. This is the smallest value of $\Gamma$ for which no oscillations occur.
For $2J < \Gamma$, the qubits exhibit slower exponential decay.
Finally, we note that \cref{eq:crit-damping} is exactly of the same form as the decoherence factor of a single qubit subjected to a {telegraph} process \cite{Neuenhahn_2009,Ayachi_2014,Cai_2020}.


\begin{thebibliography}{99}
\bibitem{com4} In the presence of degeneracy, the zeroth order eigenstates $\ketflip{\nu_k,\nu_l}^{(0)}$ are in general different from the unperturbed eigenstates $\ketflip{\nu_k,\nu_l}$.
\bibitem{com1} For a quantum many-body system,  the decoherence-free subspace can in principle contain many modes with purely imaginary eigenvalues of the Liouvillian. These eigenmodes would be unaffected by the noise and lead to nondecaying oscillations. However, we found that the quantum spin chain does not fully synchronize whenever there are several surviving modes. We note that this quantum mechanism of  noise-induced synchronization is different from the classical phenomenon discussed in Refs.~\cite{gol05,gol05a}. There,  the largest Lyapunov exponent of the system is zero in the absence of noise and turns negative in the presence of noise.
\bibitem{com3} {For the quantum $XY$ chain, stable synchronization may not  be induced by damping-type operators such as $\sigma_{x,y}$.}
\bibitem{com5} Since the decoherence-free subspace is  decoupled from the external noise, the stable synchronization condition \eqref{9} holds  independently of the value of the amplitude $\gamma$.
\bibitem{com6} The asymptotic synchronized state can only be reached via nonunitary (dissipative) dynamics. However, every single (non-averaged) realization of the noisy process is unitary (non-dissipative). As a result, only the (nonunitary) average state can exhibit synchronization.
\bibitem{com5} The synchronization rate can only reach the Lieb-Robinson limit  in a  two-spin XY chain, for which there is no stable synchronization \cite{sup}.
\end{thebibliography}
\end{document}